\documentclass[aip,pop,reprint,amssymb,amsmath,showpacs,superscriptaddress,twocolumn]{revtex4-1}

\usepackage{graphicx}
\usepackage{dcolumn}
\usepackage{bm}
\usepackage[utf8]{inputenc}
\usepackage[T1]{fontenc}
\usepackage{mathptmx}

\usepackage{color}

\newcommand{\um}{\textrm{ }\mu\textrm{m}}
\newcommand{\us}{\textrm{ }\mu\textrm{s}}

\newcommand{\vk}{v _{\rm k}}
\newcommand{\ki}{k _{\rm i}}
\newcommand{\ks}{k _{\rm s}}
\newcommand{\mi}{m _{\rm i}}
\newcommand{\NI}{n _{\rm i}}

\newcommand{\TE}{T _{\rm e}}
\newcommand{\TI}{T _{\rm i}}

\newcommand{\NE}{n _{\rm e}}
\newcommand{\LD}{\lambda _{\rm D}}

\def\vector#1{\mbox{\boldmath $#1$}}

\begin{document}

\preprint{AIP/123-QED}

\title{
	Anomalous plasma acceleration in colliding high-power laser-produced plasmas
}

\author{T. Morita}
\email{morita@aees.kyushu-u.ac.jp.}
\affiliation{Faculty of Engineering Sciences, Kyushu University, 6-1 Kasuga-Koen, Kasuga, Fukuoka 816-8580, Japan}
\author{K. Nagashima}
\affiliation{Interdisciplinary Graduate School of Engineering Sciences, Kyushu University, 6-1, Kasuga-Koen, Kasuga, Fukuoka 816-8580, Japan}
\author{M. Edamoto}
\affiliation{Interdisciplinary Graduate School of Engineering Sciences, Kyushu University, 6-1, Kasuga-Koen, Kasuga, Fukuoka 816-8580, Japan}
\author{K. Tomita}
\affiliation{Faculty of Engineering Sciences, Kyushu University, 6-1 Kasuga-Koen, Kasuga, Fukuoka 816-8580, Japan}
\author{T. Sano}
\affiliation{Institute of Laser Engineering, Osaka University, 2-6 Yamadaoka, Suita, Osaka 565-0871, Japan}
\author{Y. Itadani}
\affiliation{Interdisciplinary Graduate School of Engineering Sciences, Kyushu University, 6-1, Kasuga-Koen, Kasuga, Fukuoka 816-8580, Japan}
\author{R. Kumar}
\affiliation{Graduate School of Science, Osaka University, 1-1 Machikane-yama, Toyonaka, Osaka 560-0043, Japan}
\author{M. Ota}
\affiliation{Graduate School of Science, Osaka University, 1-1 Machikane-yama, Toyonaka, Osaka 560-0043, Japan}
\author{S. Egashira}
\affiliation{Graduate School of Science, Osaka University, 1-1 Machikane-yama, Toyonaka, Osaka 560-0043, Japan}
\author{R. Yamazaki}
\affiliation{Department of Physics and Mathematics, Aoyama Gakuin University, 5-10-1 Fuchinobe, Sagamihara, Kanagawa 252-5258, Japan}
\author{S. J. Tanaka}
\affiliation{Department of Physics and Mathematics, Aoyama Gakuin University, 5-10-1 Fuchinobe, Sagamihara, Kanagawa 252-5258, Japan}
\author{S. Tomita}
\affiliation{Department of Physics and Mathematics, Aoyama Gakuin University, 5-10-1 Fuchinobe, Sagamihara, Kanagawa 252-5258, Japan}
\author{S. Tomiya}
\affiliation{Department of Physics and Mathematics, Aoyama Gakuin University, 5-10-1 Fuchinobe, Sagamihara, Kanagawa 252-5258, Japan}
\author{H. Toda}
\affiliation{Department of Physics and Mathematics, Aoyama Gakuin University, 5-10-1 Fuchinobe, Sagamihara, Kanagawa 252-5258, Japan}
\author{I. Miyata}
\affiliation{Department of Physics and Mathematics, Aoyama Gakuin University, 5-10-1 Fuchinobe, Sagamihara, Kanagawa 252-5258, Japan}
\author{S. Kakuchi}
\affiliation{Department of Physics and Mathematics, Aoyama Gakuin University, 5-10-1 Fuchinobe, Sagamihara, Kanagawa 252-5258, Japan}
\author{S. Sei}
\affiliation{Department of Physics and Mathematics, Aoyama Gakuin University, 5-10-1 Fuchinobe, Sagamihara, Kanagawa 252-5258, Japan}
\author{N. Ishizaka}
\affiliation{Department of Physics and Mathematics, Aoyama Gakuin University, 5-10-1 Fuchinobe, Sagamihara, Kanagawa 252-5258, Japan}
\author{S. Matsukiyo}
\affiliation{Faculty of Engineering Sciences, Kyushu University, 6-1 Kasuga-Koen, Kasuga, Fukuoka 816-8580, Japan}
\author{Y. Kuramitsu}
\affiliation{Graduate School of Engineering, Osaka University, 2-1 Yamadaoka, Suita, Osaka 565-0871, Japan}
\author{Y. Ohira}
\affiliation{Department of Earth and Planetary Science, University of Tokyo, Bunkyo, Tokyo 113-0033, Japan}
\author{M. Hoshino}
\affiliation{Department of Earth and Planetary Science, University of Tokyo, Bunkyo, Tokyo 113-0033, Japan}
\author{Y. Sakawa}
\affiliation{Institute of Laser Engineering, Osaka University, 2-6 Yamadaoka, Suita, Osaka 565-0871, Japan}

\date{\today}

\begin{abstract}
	We developed an experimental platform for studying magnetic reconnection
	in an external magnetic field
	with simultaneous measurements of plasma imaging, flow velocity, 
	and magnetic-field variation.
	Here, we investigate the stagnation and acceleration 
	in counter-streaming plasmas generated by high-power laser beams.
	A plasma flow perpendicular to the initial flow directions
	is measured with laser Thomson scattering.
	The flow is, interestingly, accelerated
	toward the high-density region,
	which is opposite to the direction of the acceleration 
	by pressure gradients.
	This acceleration is possibly interpreted by the interaction of two
	magnetic field loops initially generated by Biermann battery effect,
	resulting in a magnetic reconnection forming a single field loop
	and additional acceleration by a magnetic tension force.
\end{abstract}


\maketitle



%

Magnetic fields in plasmas have significant roles in thermalization,
acceleration, and hydrodynamic turbulence in wide range of plasma
parameters, for example, 
from low-$\beta$ to high-$\beta$ conditions, 
where $\beta$ is the ratio of thermal to magnetic pressures.
Magnetic reconnection (MR) 
is the most important mechanism
in the global change of magnetic field topology and 
rapid energy transfer from the field to particles
in fusion plasmas, magnetic storms in the earth's magnetosphere, 
solar eruptions, and magnetic fields in most astrophysical contexts\cite{Zweibel2009}.
MR physics can be interpreted as a combination of 
microscopic dissipation 
and macroscopic advection in surrounding magnetized plasmas.
Previous numerical studies 
(e.g. magnetohydrodynamic\cite{Tanaka2010}
and full particle-in-cell\cite{Daughton2011} simulations) 
and observations 
(including electron and ion velocity distributions measured
by the Geotail spacecraft\cite{Nagai2011}) 
have approached 
these two mechanisms from separate viewpoints
and this makes it difficult
to investigate the physical mechanisms 
by which energy changes from across large spatial scales, and
how the reconnection rate is determined.

Laboratory experiments have an advantage in
that various plasma diagnostics can be used 
simultaneously
both for microscopic and macroscopic phenomena.
Local plasma parameters and magnetic fields have been precisely diagnosed 
in gas-discharged plasmas
such as TS-3\cite{Ono1993,Ono2011}, 
MRX\cite{Yamada1997,Ren2005}, 
and
pulse-powered devices\cite{Hare2018,Hare2017a},
with 
relatively low-beta
($\beta < 1$). 
Recently, strongly-driven MR has been studied
using laser-produced 
plasmas\cite{Totorica2017,Totorica2016,Rosenberg2015a,Rosenberg2015b,Nilson2006,Fox2012,Dong2012,Zhong2010,Nilson2008}
under strong magnetic fields generated by the interaction of high-power lasers 
with solids via the Biermann battery effect
($\partial B/\partial t \propto \nabla \TE \times \nabla \NE$)\cite{Stamper1971}.
In contrast to other laboratory experiments,
laser-produced plasmas with high temperatures and densities
enable us to investigate MR in
high-beta conditions, 
similar to those
in magnetosheath ($\beta\sim$ 0.1--10)\cite{Trenchi2008} and
in accretion disks ($\beta>10$)\cite{Balbus1998}.
However, few diagnostics are available in such small-scale plasmas,
and recent studies have focused 
on topological change in a field\cite{Rosenberg2015a,Rosenberg2015b,Nilson2006},
global plasma structure\cite{Dong2012,Zhong2010,Nilson2008},
and numerical simulations\cite{Totorica2017,Totorica2016}.
Although spatially and temporally resolved measurements 
of flow velocities, plasma parameters, and magnetic fields are required
to fully understand physical processes such as the MR rate,
no direct measurement of plasma parameters in inflow or outflow 
and magnetic fields have been performed in laser-plasma experiment.

We implemented an experimental system for studying the MR 
with simultaneous measurements of plasma imaging, flow velocity, 
and magnetic-field variation
in an external magnetic field.
In this letter, we 
investigate counter-streaming plasma interactions, stagnation, and
acceleration perpendicular to the initial flow directions.
The plasma is, interestingly, accelerated toward higher-pressure region,
showing strong temporal and spatial dependence.
This anomalous flow is possibly interpreted as the acceleration
by a magnetic tension force from a distorted magnetic field,
resulting from the MR between anti-parallel fields initially generated 
via Biermann battery effects and advected with
expanding plasma\cite{D.D.RyutovN.L.KuglandM.C.LevyC.PlechatyJ.S.Ross2013}.

%
\begin{figure}
\begin{center}
\includegraphics[width=0.99\linewidth]{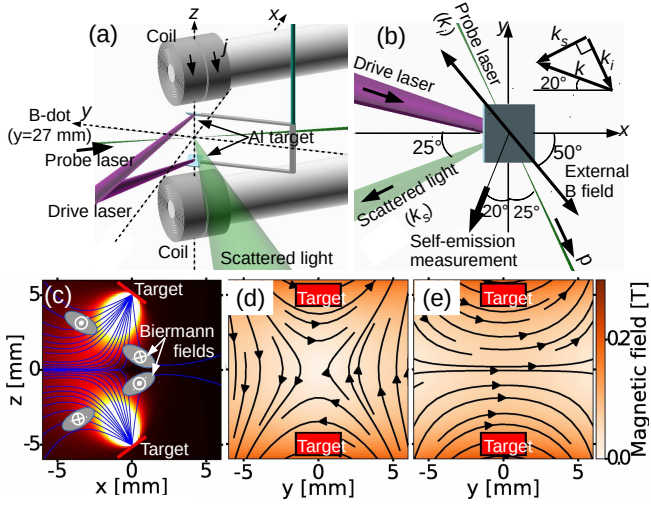}
\caption{\label{fig:setup} 
	(a) Schematic
	of the experimental setup with two electromagnetic
	coils and two targets. 
	The drive laser and probe laser for TS measurement
	are also shown.
	(b) The top view of the setup. The probe laser ($\vector{\ki}$) and 
	the direction of the scattered light
	($\vector{\ks}$) determine 
	the $\vector{k}$-vector.
	(c) Calculated electron streams from 
	tilted targets, as well as the Biermann field in the $x$-$z$ plane.
	The external magnetic fields are
	generated by driving two coils 
	in the $y$-$z$ plane,
	(d) in opposite directions (anti-parallel)
	and
	(e) in the same direction (parallel).
}
\end{center}
\end{figure}

The experiment was conducted with Gekko-XII laser beams at the Institute of
Laser Engineering, Osaka University.
Figure \ref{fig:setup}(a) shows the schematic of the experimental setup.
Two laser beams (drive laser: output energy 600 J 
in a 1.3 ns Gaussian pulse with wavelength 1053 nm)
irradiate two aluminum foils (with thickness 200 $\um$ 
and separation 10 mm)
with focal spot diameters of $\sim$300 $\um$,
to
produce two plasma flows between them.
The targets are tilted 30$^\circ$ from 
the horizontal plane to produce 
tilted flows in the $x$-$z$ plane 
as shown in Fig. \ref{fig:setup}(c).
Plasma flow velocity was measured with laser Thomson scattering (TS) in
the center of the targets.
A probe laser 
(Nd:YAG laser, wavelength 532 nm, energy 370 mJ in 10 ns,
wavenumber vector $\vector{\ki}$) 
penetrated
the plasma 
in
the horizontal plane ($z=0$)
at an angle of
65$^\circ$ from the $x$-axis,
and the scattered light ($\vector{\ks}$) was collected 
90$^\circ$ from the incident direction, forming the measurement wavenumber,
$\vector{k}=\vector{\ks}-\vector{\ki}$ as shown in Fig. \ref{fig:setup}(b)
(20$^\circ$ rotated from the $x$-axis).
The scattered light was guided to a high-resolution ($\sim$14 pm) spectrometer
with triple gratings\cite{Tomita2017} and recorded with
an intensified 
charge-coupled
device (ICCD) 
with an exposure time of 5 ns.
The expanding plasmas 
were imaged
with a framing camera at the wavelength of 450 nm with an interference
filter with the bandwidth of 10 nm (FWHM).
In addition, a magnetic-field variation ($\Delta B_z$) 
at $y=27$ mm was measured with a B-dot probe
consisting of two inversely directed loops. Most of the electromagnetic noise
is reduced by taking difference between the voltages from these loops.

As studied in many experiments with laser-produced counter-streaming
plasmas\cite{Huntington2015,Sakawa2016,Fox2013,Ross2012b,Yuan2013,Kugland2012a,Park2012,Kuramitsu2011,Morita2010b},
the magnetic field originated around the laser spot 
(by the
Biermann battery effect) is
frozen into the ablation plasma and 
is advected along electron stream lines\cite{D.D.RyutovN.L.KuglandM.C.LevyC.PlechatyJ.S.Ross2013} because of the large magnetic Reynolds number.
Note that the typical time scale of the Bierman battery 
effect $\tau_{\rm Bir}$ is comparable to the laser pulse duration $\sim$1 ns, 
which is much smaller than typical time scale of plasma dynamics
$\tau_{\rm dyn} \sim 40$ ns in the present study.
The electron stream lines are estimated and plotted 
for 
tilted targets [Fig. \ref{fig:setup}(c)].
Here, the electron flux is evaluated as 
$\phi(r,\theta) \propto \exp[-K(1-\cos{\theta})]/r^2$, where
$r$ and $\theta$ represent the distance from the laser spot
and the angle from the normal direction, respectively, 
and $K$ characterizes the divergence as discussed in Ref. \onlinecite{D.D.RyutovN.L.KuglandM.C.LevyC.PlechatyJ.S.Ross2013}.
When the targets are parallel\cite{D.D.RyutovN.L.KuglandM.C.LevyC.PlechatyJ.S.Ross2013},
the toroidal magnetic fields from the top and bottom targets encounter
each other in the mid-plane. 
On the other hand,
the toroidal fields interact only around the mid-point
in the tilted flows as shown in Fig. \ref{fig:setup}(c).

In addition, an external magnetic field was applied with two 20-turn coils
located at $z=\pm17$ mm along the axis 
at
40$^\circ$ from the $y$-axis,
as shown 
with an arrow in Fig. \ref{fig:setup}(b).
They are driven separately by a pulse-powered device\cite{Edamoto2018}.
An anti-parallel [Fig. \ref{fig:setup}(d)] field enhances
the $y$-component of the Biermann fields ($B_{{\rm Bir},y}$) both 
for
$z>0$ and $z<0$,
while a parallel [Fig. \ref{fig:setup}(e)] field weakens
$B_{{\rm Bir},y}$ 
for
$z<0$.
In the present experiment, a magnetic field strength of 0.1 T
at the target positions was achieved by applying 200 V 
on a 6-mF capacitor\cite{Morita2017a} for each coil, 
producing a more or less
constant field during $\sim$100 $\us$.

%
\begin{figure}
\begin{center}
\includegraphics[width=0.99\linewidth]{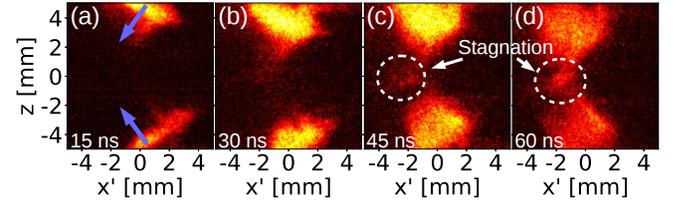}
\caption{\label{fig:emission} 
	Self-emission from laser-produced plasmas at the wavelength of 
	450 nm at $t=$ (a) 15 ns, (b) 30 ns, (c) 45 ns, and
	(d) 60 ns,
	taken with a framing camera.
}
\end{center}
\end{figure}

\begin{figure}
\begin{center}
\includegraphics[width=0.99\linewidth]{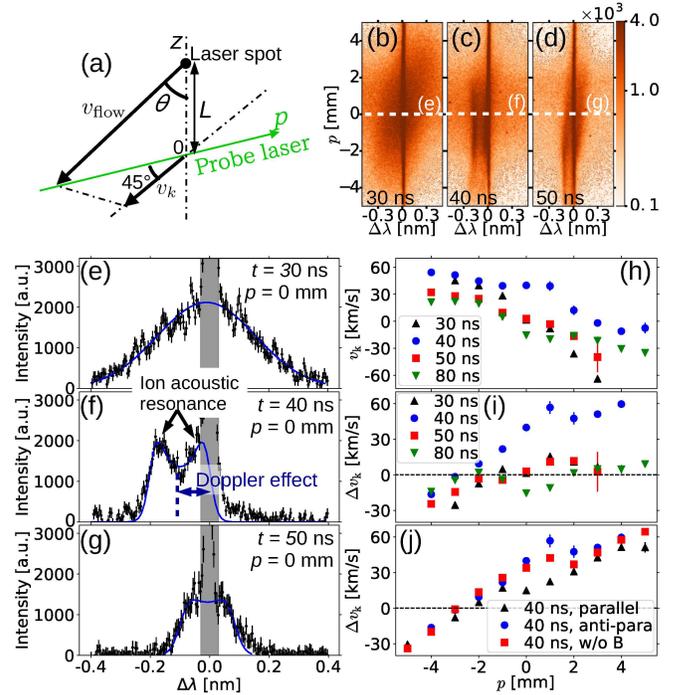}
\caption{\label{fig:TS} 
	(a) The relation between the flow velocity 
	from the upper target ($v_{\rm flow}$)
	and the measured velocity ($v_{k}$).
	TS spectra 
	are given at (b) 30 ns, (c) 40 ns, and (d) 50 ns
	along the probe laser axis. The origin $p=0$ corresponds
	to the midpoint between two laser spots.
	The line-out plots are given
	at (e) 30 ns, (f) 40 ns,
	and (g) 50 ns for $p=0$ mm.
	The solid lines show the best-fit results.
	(h) shows $v_{k}$ and 
	(i) shows the velocity difference $\Delta\vk$ measured for 
	different delay times
	with an anti-parallel external field, and
	(j) shows $\Delta\vk$ in different external field configurations.
}
\end{center}
\end{figure}

Figure \ref{fig:emission} shows the time-evolution of the plasma
after the laser irradiation 
on
two aluminum foils 
measured 
at 
20$^\circ$ from the $y$-axis, as shown 
in Fig. \ref{fig:setup}(b) 
(the horizontal axis $x'$ is the axis 20$^\circ$ rotated
from $x$) with the anti-parallel external field [Fig. \ref{fig:setup}(d)].
Note that the external field does not affect 
the initial super-Alfv\'enic flows.
These are flows for which the Alfv\'en Mach number
$M_{\rm A}=v/v_{\rm A}\gtrsim $ 51, 
and hence the magnetic pressure $P_M$
is much smaller than the ram pressure $\rho v^2$
($P_M/\rho v^2 = 1/2M_{\rm A}^2\lesssim 2\times10^{-4}$),
where $v\gtrsim125$ km/s is the flow velocity
estimated at $t<40$ ns,
$v_{\rm A}=2.5\pm 0.4$ km/s is the Alfv\'en velocity
in an external magnetic field of 0.1 T 
[two different geometries are shown in Figs. \ref{fig:setup}(d) and \ref{fig:setup}(e)]
with ion density
$\NI=(2.9\pm 0.5)\times10^{22}$ m$^{-3}$
estimated from TS spectrum
at $t=40$ ns;
this will be discussed later in Fig. \ref{fig:TS}(f).
As the plasmas expand from top and bottom, two flows interact
to form a dense plasma which is observed 
in the mid-plane at $t \gtrsim 45$ ns.
	Note that this dense plasma is reproducibly observed in different laser shots (not shown)
at $-4$ mm $<x'<-1$ mm.

Figures \ref{fig:TS}(b)--\ref{fig:TS}(d) 
show the TS spectra 
at $t=$ 30, 40, and 50 ns, respectively, 
with the anti-parallel external magnetic field 
[Fig. \ref{fig:setup}(d)].
The vertical axis shows the position along the probe laser 
[$p$-axis: $(x,y)=(p\cos 65^\circ,\,p\sin 65^\circ)$] 
and the horizontal axis represents the wavelength difference
$\Delta\lambda=\lambda-\lambda_0$, where $\lambda_0$ is the incident
laser wavelength. 
A bright stray light is observed at $\Delta\lambda = 0$.
As illustrated in Fig. \ref{fig:setup}(a), the plasma parameters
parallel to $\vector{k}$ 
are 
obtained in this measurement.
Figures \ref{fig:TS}(e)--\ref{fig:TS}(g) 
show the line-out plots 
for $p=0$ mm at $t=30$, 40, and 50 ns, respectively.
Two peaks of the ion acoustic resonance
are seen in Fig. \ref{fig:TS}(f),
where one of them overlaps with the stray light (the shaded area).
The solid lines in Figs. \ref{fig:TS}(f) and \ref{fig:TS}(g)
show the best-fit results with 
a theoretical function\cite{PlasmaScattering.Froula}
convoluted with the spectral resolution\cite{Morita2013a} 
(estimated from Rayleigh scattering),
where the charge state $Z$ is self-consistently obtained from 
a collisional radiative model with FLYCHK code\cite{H.-K.Chung2005}.
Note that this model is valid, in general, in highly ionized and
moderate density plasmas,
as observed in the present experiment,
between high-density (local thermal equilibrium) and
low-density (coronal equilibrium) limits,
where excited populations are determined by
collisional and radiative processes.
At $t=40$ ns, the fitting gives 
electron and ion temperatures of
$\TE=92\pm35$ eV and $\TI=75\pm14$ eV, 
respectively,
electron density
$\NE=(2.6\pm0.4)\times10^{23}$ m$^{-3}$,
and average charge state
$Z=9.0 \pm 1.0$.
At $t=50$ ns,
the corresponding values are
$\TE=74\pm23$ eV, $\TI=87\pm17$ eV, 
$\NE=(1.6\pm0.2)\times10^{23}$ m$^{-3}$,
and $Z=8.1 \pm 0.8$.
In general, ion-feature is 
in collective regime when 
$\alpha \gtrsim \left(Z\TE/3\TI-1\right)^{-1/2}$\cite{PlasmaScattering.Froula},
where $\alpha = 1/k\LD$ and $\LD$ is the Debye length.
Although the ion-acoustic wave is weakly damped at $t=40$ ns 
($\alpha=0.43\pm0.09$ is slightly under $\left(Z\TE/3\TI-1\right)^{-1/2}=0.62\pm0.19$),
the ion-acoustic resonance is clearly seen
for $-4$ mm $<p<$ 2 mm in Fig. \ref{fig:TS}(c),
and $-4$ mm $<p<$ 0 mm in Fig. \ref{fig:TS}(d). 
On the other hand, no ion-acoustic resonance is observed 
at $t=30$ ns [Fig. \ref{fig:TS}(e)].
Assuming the non-collective ion-feature of the ion thermal distribution as Maxwellian,
a large ion temperature of $\TI=1.2\pm0.1$ keV is obtained
as shown 
by
the solid line in Fig. \ref{fig:TS}(e),
indicating small $Z\TE/\TI$
and strong damping of ion-acoustic waves.

The flow velocity along $\vector{k}$ ($\vk$) can be estimated
from the deviation of the spectrum from $\Delta\lambda=0$ 
(Doppler effect)
in both the collective and non-collective regimes.
The standard errors are estimated by non-linear least square fitting 
and are smaller than the marks in most of data points 
in Figs. \ref{fig:TS}(h)--\ref{fig:TS}(j).
Although there are systematic errors, namely
the pointing of the probe laser ($<0.1$ mm) and
timing jitter of the detector ($\sim35$ ps),
both are ignorable when determining $\vk$.
Early 
and late 
in time,
as shown in Figs. \ref{fig:TS}(b) and \ref{fig:TS}(e) 
($t=30$ ns),
and \ref{fig:TS}(d) and \ref{fig:TS}(g) ($t=50$ ns),
respectively,
the flow velocity is small 
for 
$p\sim0$ mm
($\vk=4.1\pm1.4$ km/s and $2.8\pm0.9$ km/s 
from Figs. \ref{fig:TS}(e) and \ref{fig:TS}(g), respectively).
In contrast,
the TS spectrum obtained at $t=40$ ns 
[Figs. \ref{fig:TS}(c) and 
\ref{fig:TS}(f)],
is blue-shifted at $p < 2$ mm 
which means that the plasma flows into the positive $\vector{k}$-direction
($\vk=40\pm3$ km/s at $p=0$).
This tendency is seen in Fig. \ref{fig:TS}(h),
where $\vk$ is plotted as a function of
$p$ measured at $t=30$, 40, 50, and 80 ns.

As illustrated in Fig. \ref{fig:TS}(a),
the plasma flow in the case of free streaming can be estimated as
$v_{\rm k,free} = -v_{\rm flow}\sin\theta$$\cos 45 ^\circ$,
where $v_{\rm flow} = (L^2+p^2)^{1/2}/t$ is the flow velocity 
from the laser spot to the measured position,
$\theta$ is the angle defined in Fig. \ref{fig:TS}(a),
$L$ is the half distance between two laser spots,
and $t$ is the time from the drive laser timing.
$v_{\rm k,free}$ is a good estimation for the flow velocity 
even in counter-streaming plasmas
as observed in similar experimental configurations 
in Refs. \onlinecite{Ross2012b} and \onlinecite{Morita2013a},
in which slow-down of each flow is only 10--20 \%.
Figure \ref{fig:TS}(i) shows the difference between 
the measured velocity $\vk$ and 
the calculated velocity $v_{\rm k,free}$,
$\Delta \vk = \vk - v_{\rm k,free}$.
Early and late in time at $t=30$, 50, and 80 ns,
although 
$\Delta\vk \sim 0$
near the center $p\sim0$,
$|\vk|<|v_{\rm k,free}|$ far from the center
due to the slow-down of the plasma flow.
Note that the slow-down effect shows
$\Delta\vk > 0$ and $\Delta\vk<0$ for $p>0$ and $p<0$, respectively.
While this slow-down effect on $\Delta \vk$ is less than 30 km/s,
the velocity difference of $\sim$60 km/s at $t=40$ ns 
is much larger than that
at $p \sim 1$  mm,
meaning that another acceleration occurs only at this time.
Figure \ref{fig:TS}(j) shows $\Delta\vk$
under
three different external field conditions:
with an anti-parallel field [Fig. \ref{fig:setup}(d)], 
with a parallel field [Fig. \ref{fig:setup}(e)], 
and without the external field.
The measured velocity with an anti-parallel field 
($\Delta\vk = 57\pm 5$ km/s)
is larger than that without the external field 
($\Delta\vk = 42\pm 1$ km/s)
at $p=1$ mm.
All TS data with different times and external field conditions 
are taken in different laser shots,
indicating that the acceleration along $\vector{k}$ at $t=40$ ns
is reproducible as expressed in Fig. \ref{fig:TS}(j).

\begin{figure}
\begin{center}
\includegraphics[width=0.99\linewidth]{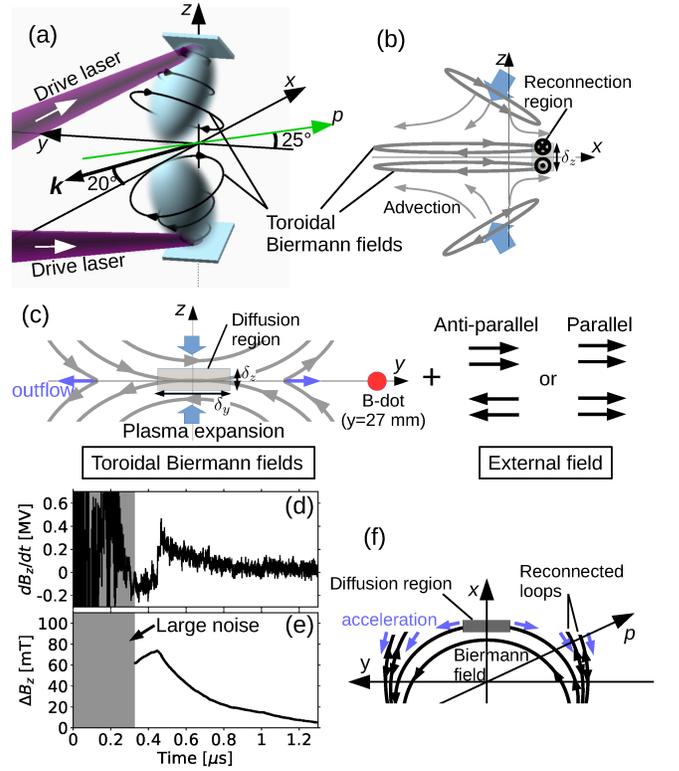}
\caption{\label{fig:Bfield} 
	(a) Three-dimensional and (b) two-dimensional (in the $x$-$z$ plane)
	cartoons of the toroidal Biermann fields in the top and bottom plasmas.
	(c) Two toroidal fields interact in the mid-plane as shown in 
	a two-dimensional view (in the $y$-$z$ plane) 
	with two different configurations of external magnetic fields.
	(d) Time-derivative of the magnetic-field variation ($dB_z/dt$)
	and (e) the calculated magnetic-field deviation $\Delta B_z$ 
	from the constant field later in time ($t>2$ $\us$)
	measured at $y=27$ mm.
	(f) The schematic view of the field lines in the $x$-$y$ plane after MR.
}
\end{center}
\end{figure}

Two possibilities 
may explain this anomalous acceleration:
(1) a pressure increase 
in 
the mid-plane resulting from the collision
of two plasma flows, and (2) MR between toroidal fields
generated via the Biermann battery effect.
For the former case, assuming 100\% energy conversion 
from initial kinetic energy to thermal energy in the mid-plane,
and the resultant kinetic energy accelerated by the thermal pressure,
a value of $v_{\rm k} < 125$ km/s 
(estimated with $L=5$ mm and $t=40$ ns) 
can be explained.
It might be possible that the hydrodynamic jet is driven from the stagnation point
where
two plasmas collide with each other forming higher pressure at $x\sim -2$ mm as shown in 
Figs. \ref{fig:emission}(c) and \ref{fig:emission}(d).
This flow (along pressure gradients) 
has negative $\Delta\vk$ at $x\sim0$,
while we observe the positive $\Delta\vk$ as shown in
Figs. \ref{fig:TS}(i) and \ref{fig:TS}(j).
This indicates that the observed flow in the $\vector{k}$-direction 
[see Figs. \ref{fig:TS}(i) and \ref{fig:TS}(j)] is not
explained by the pure hydrodynamics,
but is rather interpreted as the acceleration by the magnetic field.
In this case,
the Biermann fields become flat structures as they collide with each other
and stagnate in the mid-plane
as demonstrated in Ref. \onlinecite{Li2013}
even in a tilted geometry similar to the present experiment.
These fields [Fig. \ref{fig:Bfield}(a)] can reconnect with each other
as illustrated in Fig. \ref{fig:Bfield}(b),
forming a bent loop 
as shown in Ref. \onlinecite{Li2016a}.
The thickness of the magnetic diffusion region 
for 
the MR is comparable to the ion inertial length\cite{Zweibel2009},
$\delta_z \sim c/\omega_{\rm pi} = 0.77 \pm0.16$ mm,
calculated with the parameters
obtained from the TS spectrum at $t=40$ ns 
[Fig. \ref{fig:TS}(f)].
This small MR region 
indicates quasi-two-dimensional MR
in the $y$-$z$ plane because $B_x$ is small
($B_{{\rm Bir},x}/B_{{\rm Bir},y}\sim \delta_x / \delta_y \sim 0.29$,
where 
$\delta_x \sim \delta_z$ and
$\delta_y \sim [R^2-(R-\delta_x)^2]^{1/2}\sim2.7$ mm
are the lengths of the diffusion region in the $x$ and $y$-directions
assuming
a large curvature radius of the Biermann field
$R = L=5$ mm),
and the MR should initially produce outflows predominantly 
in the $\pm y$-directions
as shown in Fig. \ref{fig:Bfield}(c).

Here, the magnetic Reynolds number 
in the present experiment is large: 
$R_{\rm M} = Lv/D_{\rm M}=(2.2\pm 1.1)\times10^2$,
where $L\sim5$ mm is the scale length, 
$v\sim$ 125 km/s at $t=$ 40 ns is the vertical flow velocity,
$D_{\rm M} = \nu_{\rm ei}(c/\omega_{\rm pe})^2 = 2.8 \pm 1.4$ m$^2$/s
is the magnetic diffusivity,
$\nu_{\rm ei}=(2.6\pm 1.4)\times10^{10}$ s$^{-1}$
is the electron-ion collision frequency\cite{Chena},
and $c/\omega_{\rm pe}=10\pm 2$ $\um$ is the electron skin depth.
Typically, ideal magnetohydrodynamics (MHD) 
can be applied\cite{Ryutov2000} to plasmas
with $R_{\rm M}$ larger than 10. 
In addition, larger $R_{\rm M}$ ($\sim$$10^2$--$10^3$) enables us
to investigate MR with a current sheath
much thinner than typical plasma scale\cite{Ryutov2000}. 
The magnetic force acting on the unit volume of the plasma is expressed as
\begin{align}
	\vector{f} &= \vector{j}\times \vector{B} = (\vector{\nabla}\times \vector{B})\times \vector{B}/\mu_0 \nonumber \\ 
	&= -\nabla({B^2}/{2\mu_0}) + {\vector{B}\cdot\nabla\vector{B}}/{\mu_0} \nonumber \\
	&= -\nabla_\perp(B^2/2\mu_0) - B^2\hat{r_c}/\mu_0r_c = \vector{f}_p + \vector{f}_t, \nonumber 
\end{align}
where $\mu_0$ is the vacuum permeability; $\nabla_\parallel$ and $\nabla_\perp$ are the gradients 
parallel and perpendicular to the magnetic field, respectively; and
$r_c$ and $\hat{r_c}$ are the curvature radius and the unit vector
along the radial direction, respectively.
The first term $\vector{f}_p$ represents 
the magnetic pressure gradient force 
perpendicular to the field line, and the second term 
$\vector{f}_t$ expresses
the magnetic tension force when the field line curves.
While $\vector{f}_p$ acts isotropically vertical to the field, 
$\vector{f}_t$ accelerates the plasma 
towards $\hat{r_c}$.
This tension force acts in the $\pm y$-direction 
near the diffusion region
generating an outflow from MR,
which is quasi-two-dimensional
[see the left figure in Fig. \ref{fig:Bfield}(c)].
The plasma is accelerated 
on a typical scale of $l$ 
in the acceleration time of $\Delta t \sim (2l/a)^{1/2}$
where $a=f_t/\mi\NI$ is the acceleration. The resultant 
plasma velocity becomes
\begin{align}
	v_{\rm acc} &\sim |a|\Delta t = (2la)^{1/2} 
	= B(2/\mu_0\mi\NI)^{1/2}(l/r_c)^{1/2} \nonumber \\
	&\sim B(2/\mu_0\mi\NI)^{1/2} \sim v_{\rm A}, \label{eq:vacc}
\end{align}
where 
$\NI=(2.9\pm 0.5) \times 10^{22}$ m$^{-3}$,
and the acceleration distance $l$ is comparable to the curvature radius
$l \sim r_c$.
The time-derivative of the magnetic-field variation in the $z$-component 
($dB_z/dt$)
and the magnetic-field deviation [$\Delta B_z=-\int_{2\us}^t (dBz/d\tau) d\tau$] from a constant
field strength later in time ($>2$ $\us$), without the external magnetic field,
are shown in Figs. \ref{fig:Bfield}(d) and \ref{fig:Bfield}(e), respectively.
Although the magnetic field early in time is difficult to infer 
due to large electromagnetic noise, a positive $\Delta B_z$ is detected
at $t\sim450$ ns, suggesting a magnetized plasma flow with
the velocity of $\sim$66 km/s [$=27$ mm$/(450-40)$ ns]
assuming MR at $t=40$ ns,
which is comparable to the Alfv\'en velocity
with $B\sim 2.7$ T and $\NI=2.9 \times 10^{22}$ m$^{-3}$
at the acceleration site.

As the reconnected loop expands in $\pm y$, 
$\vector{f}_t$ also accelerates the plasma in the $-x$-direction
because
the $\hat{r_c}$ of a reconnected field loop is in the $-x$-direction
as shown in the $x$-$z$ plane [Fig. \ref{fig:Bfield}(f)] at $p>0$.
The measured velocity $\vk\sim57$ km/s can be explained with the
field strength of $\sim$2.3 T which is slightly smaller than
the strength of $\sim$2.7 T estimated from the velocity in the $y$-direction.
The initial curvature radius $r_c$ ($y$-direction) is formed as a result of MR, and 
the curvature in the $x$-direction would be comparable to or slightly larger than that
as the field expand in the $\pm y$-directions;
$r_c\gtrsim \delta_z \sim c/\omega_{\rm pi}\sim0.77$ mm.
Note that 
the long diffusion time 
$\tau_{\rm M}$ of the magnetic field 
in the typical plasma size of $L$
($\tau_{\rm M} = L^2/D_{\rm M} 
= (L/v)R_M \sim \tau_{\rm dyn}R_M = 8.8\pm4.4$ $\us \gg \tau_{dyn}
\sim 40$ ns
for $L=5$ mm)
indicates that 
only the surface of the expanding plasma 
is magnetized, and 
MR occurs only when the plasmas collide 
with 
each other at $t\sim40$ ns.
The acceleration due to the magnetic tension force acts on the plasma
after the MR,
and this is why
the TS measures the velocity increase only at $t=40$ ns.

From Eq. (\ref{eq:vacc}), the increment in the
velocity from 42 km/s (only the Biermann fields) to 
57 km/s (now with the anti-parallel external field) 
can be explained by a $57/42\sim1.4$ times stronger 
magnetic field,
or increment in the field from 1.6 to 2.3 T.
Even though the initial external field strength of 0.1 T is weak 
($M_{\rm A} > 52$ at $t<40$ ns) 
and does not affect the initial plasma streams,
it can be piled up as the plasma expands because of high conductivity,
which 
has already been
seen in previous experimental 
studies\cite{Rosenberg2015,Rosenberg2015a,Rosenberg2015b,G.Fiksel,Fox2012}
by a factor of four,
meaning that both parallel and anti-parallel external fields
become comparable to the Biermann fields 
in 
the interaction region.
As investigated in 
Ref. \onlinecite{D.D.RyutovN.L.KuglandM.C.LevyC.PlechatyJ.S.Ross2013} 
considering the electron stream,
the 
Biermann field of $\sim$50 T 
(measured in Ref. \onlinecite{Li2006} with a comparable laser intensity)
decreases to $\sim$4 T as it is advected to 5 mm\cite{D.D.RyutovN.L.KuglandM.C.LevyC.PlechatyJ.S.Ross2013},
and the reconnected field is even weaker;
this agrees well with our estimation of 
$\sim$2.7 T for the acceleration in the $y$-direction
and $\sim$1.6--2.3 T in the $x$-direction after MR with and without 
anti-parallel external magnetic field.

Another possible interpretation for the external magnetic field dependence
comes from the $x$-component of the external magnetic field.
The anti-parallel external field ($\pm B_{{\rm ext},x}$, $\pm B_{{\rm ext},y}$)
and the Biermann field (0, $\pm B_{{\rm Bir},y}$) make
another anti-parallel field of 
($\pm B_{{\rm ext},x}$, $\pm B_{{\rm ext},y}\pm B_{{\rm Bir},y}$)
on a different plane rotated around $z$,
where positive and negative signs show the magnetic field in
$z>0$ and $z<0$, respectively.
On the other hand, the parallel external field produces
the composite field of 
($B_{{\rm ext},x}$, $B_{{\rm ext},y}\pm B_{{\rm Bir},y}$),
meaning that the asymmetric MR between $B_{{\rm ext},y}\pm B_{{\rm Bir},y}$
[see Fig. \ref{fig:Bfield}(c)]
occurs in a guide field of $B_{{\rm ext},x}$.
This asymmetric MR decreases the velocity in the mid-plane,
resulting in the slowest velocity ($\Delta\vk = 22\pm 1$ km/s).

In summary, we have investigated the interaction between
laser-produced counter-streaming plasmas, stagnation, and acceleration,
with newly implemented experimental platform with 
self-emission imaging, laser Thomson scattering, and a B-dot probe.
The plasma acceleration perpendicular to the initial flow directions
is detected.
This flow is, interestingly, opposite to the acceleration 
due to pressure gradients, suggesting other acceleration mechanism.
Here, we proposed the acceleration via magnetic tension force
in a bent field loop resulting from a MR
between two field loops
initially generated by a Biermann battery effect and advected
with expanding plasmas.
Even though further investigation and comparison would be required,
the flow velocity changes depending on the external magnetic field
direction, indicating that the external field changes the strength
and structure of the reconnection field.
Interestingly, the acceleration quasi-perpendicular to 
the $y$-$z$ plane indicates that the global field structure
affects the microscopic plasma flow despite 
the MR occurring in the quasi-two-dimensional plane.
Though there have been no direct observation of this anomalous acceleration,
it could be detected,
for example, in the turbulent magnetosheath or solar coronal loops,
with future high-resolution measurements.
Further experimental investigations
combining multi-dimensional velocity measurements
with local magnetic field measurements
will promote greater understanding of not only the physical process 
but also the plasma dynamics of the MR.

The authors would like to acknowledge the dedicated technical support 
of the staff at the Gekko-XII facility for the laser operation, 
target fabrication, and plasma diagnostics.
We would also like to thank K. Shibata, N. Yamamoto, A. Asai,
and S. Zenitani for helpful comments and valuable discussions.
This research was partially supported by JSPS KAKENHI grant numbers
18H01232, 17H06202, 15H02154, 17K14876, and 17H18270, 
and by the joint research project of 
Institute of Laser Engineering, Osaka University.


%
\end{document}